\documentclass[useAMS,usenatbib,referee]{mn2e}
\usepackage[dvips]{epsfig}
\usepackage[dvips]{graphics,color}

\title[Feeding Versus Feedback in Mrk\,766]{Feeding Versus Feedback in AGNs from Near-Infrared IFU Observations: The Case of Mrk\,766} 
\author[Sch\"onell, A. J. J\'unior]{Sch\"onell, A. J. J\'unior$^{1}$\thanks{E-mail:
juniorfisicoo@gmail.com}, Rogemar A. Riffel$^{1}$, Thaisa Storchi-Bergmann$^{2}$, \newauthor Claudia Winge$^{3}$\\
$^{1}$Universidade Federal de Santa Maria, Departamento de F\'isica, Centro de Ci\^encias Naturais e Exatas, 97105-900,\\
Santa Maria, RS, Brazil \\
$^{2}$Instituto de F\'isica, Universidade Federal do Rio Grande do Sul, Av. Bento Goncalves 9500, 91501-970\\ 
Porto Alegre, RS, Brazil \\
$^{3}$Gemini Observatory Southern Operations Center c/o AURA, Casilla 603 La Serena, Chile }

\begin{document}

\date{11/01/2013}

\pagerange{\pageref{firstpage}--\pageref{lastpage}} \pubyear{2013}

\maketitle

\label{firstpage}

\begin{abstract}

We have mapped the emission-line flux distributions and ratios as well as the
gaseous kinematics of the inner 450 pc radius of the Seyfert 1 galaxy Mrk\,766
using integral field near-IR J- and K$_l$-band spectra obtained with the Gemini {\sc nifs} at a spatial resolution of 60 pc and velocity resolution 
of 40 km\,s$^{-1}$. Emission-line flux distributions in ionized and molecular gas extend up to $\approx$ 300 pc from
the nucleus. Coronal [S\,{\sc ix}]$\lambda1.2523\mu$m line emission is
resolved, being extended up to 150 pc from the nucleus. 
At the highest flux levels, the [Fe\,{\sc ii}]$\lambda$1.257$\mu$m line emission is most
extended to the south-east, where a radio jet has been observed.
The emission-line ratios [Fe\,{\sc ii}]$\lambda1.2570\mu$m/$Pa\beta$ and $H_{2}\lambda 2.1218 \mu$m/Br$\gamma$ show a mixture of Starburst and Seyfert excitation; the Seyfert excitation dominates at the nucleus, to the north-west and in an arc-shaped region between 0\farcs2 and 0\farcs6 to the south-east at the location of the  radio jet. A contribution from shocks at this location is supported by enhanced [Fe\,{\sc ii}]/[P\,{\sc ii}] line ratios and  increased [Fe\,{\sc ii}] velocity dispersion.
The gas velocity field is dominated by rotation that is more compact for H$_2$ than for Pa$\beta$, indicating that the molecular gas has a colder kinematics and is located in the galaxy plane. There is about 10$^{3}$ M$_{\odot}$ of hot H$_2$, implying $\approx$ 10$^{9}$ M$_{\odot}$ of cold molecular gas. 
At the location of the radio jet, we observe an increase in the [Fe\,{\sc ii}] velocity dispersion (150 km s$^{-1}$), as well as both blueshift and redshifts in the channel maps, supporting the presence of an outflow there. The ionized gas mass outflow rate is estimated to be $\approx$ 10 M$_{\odot}$ yr$^{-1}$, and the power of the outflow $\approx$ 0.08 L$_{bol}$.


\end{abstract}


\section{Introduction}

The study of the extended emission in the Narrow-Line Region (NLR) around nearby Active Galactic Nuclei (AGN) allows the investigation of both the AGN feeding -- via gas inflows \citep[e.g.][]{sbe,fa,sm,ms2009,d2009} and feedback --  via the interaction of the AGN radiation and mass outflow with the circumnuclear gas, affecting its kinematics and excitation \citep[e.g.][]{fw,fe,fe2,ce,ce2,ce3,ck,hh,vc,sk,vg,we,msea}. 

Most studies on the feeding and feedback mechanisms of Active Galactic Nuclei (AGN) presently available in the literature are based on optical observations, which are affected by dust obscuration, 
a problem that can be softened by the use of infrared observations. Another advantage of  infrared spectral region is that, besides observing ionized gas emission, we can also observe emission from molecular gas (H$_2$). Our group, {\it {\sc agnifs}} (for AGN Integral Field Spectroscopy), has been developing a project to map both the feeding and feedback in nearby AGN using near-infrared integral field spectroscopic observations mostly with the instrument {\sc nifs} at the Gemini North Telescope. 
The main findings of our group so far have been that the molecular gas -- traced by K-band $H_2$ emission, and the ionized gas traced by H{\sc i} recombination lines and [Fe\,{\sc ii}] emission, present distinct flux distributions and kinematics. Usually the $H_2$ emitting gas is restricted to the plane of the galaxy, while the ionized gas extends also to high  latitudes and is associated with the radio emission, when present \citep{re,re3,re4,rsn,rs,sbe2,sbe3}. The $H_2$ kinematics is usually dominated by rotation, including in some cases, streaming motions towards the nucleus, while the kinematics of the ionized gas, and in particular of the [Fe\,{\sc ii}] emitting gas, shows also, in many cases, a strong outflowing component associated with radio jets from the AGN. Similar results have been found using the Spectrograph for INtegral Field Observations in the Near Infrared (SINFONI) at the Very Large Telescope (VLT). \citet{d2009} found molecular gas inflows towards the nucleus of NGC\,1097 and \citet{ms2009} mapped similar H$_2$  inflows feeding and obscuring the active nucleus of NGC\,1068, while  \citet{msea} mapped outflows in ionized gas around 7 active galatic nuclei.  

In this work, we present the gaseous distribution and kinematics of the inner 450\,pc radius of the Narrow-Line Seyfert 1 galaxy Mrk\,766 (NGC\,4253) a barred spiral galaxy (SBa), located at a distance of 60.6\,Mpc, for which 1$^{\prime\prime}$ corresponds to 294\,pc at the galaxy. The HST images of this galaxy show some irregular dust filaments around the nucleus \citep{ma}. \citet{kk} show that the radio source appears to be extended to south-east in PA $\approx$ 150$^{\circ}$ (on a scale of $\approx$ 1$\farcs$). The optical emission is extended beyond the radio structure \citep{gdp96}. The NIR spectrum is well described by \citet{ardila05}, showing a large number of permited lines of H\,{\sc i}, He\,{\sc i}, He\,{\sc ii} and Fe\,{\sc ii}, and by forbidden lines of [S\,{\sc ii}], [S\,{\sc iii}] and [Fe\,{\sc ii}]. High ionization lines like [Si\,{\sc ix}], [Si\,{\sc x}], [S\,{\sc ix}] and [Mg\,{\sc viii}] are also observed. The X-rays observations of this galaxy show that it is a strong variable source, with evidences of the amplitude being larger at $\approx$ 2\,kev. The mass of the supermassive black hole has been accurately measured via reverberation mapping by \citet{bentz2009}, resulting in a mass of 1.76$^{+1.56}_{-1.40}\,\times$10$^6$\,M$_{\odot}$. There is no CO observations for this galaxy in the literature.
 
Mrk\,766 was selected for this study because: (i) it presents strong near-IR emission lines \citep[e.g.][]{ardila05}, allowing the mapping of the gaseous distribution and kinematics; and (ii) it has radio emission, allowing the investigation of the role of the radio jet \citep{kk} in the gas excitation and kinematics.
This paper is organized as follows: In Sec. 2 we describe the observations and data reduction procedures. The results are presented in Sec. 3 and discussed in Sec. 4. We present our conclusions in Sec. 5.

\section{Observations and data reduction}

The observations of Mrk\,766 were obtained using the Gemini Near Infrared Integral Field Spectrograph ({\sc nifs} - \citet{mcg}) operating with the Gemini North Adaptive Optics system ALTAIR in June 2010 under the programme GN-2010A-Q-42,
following the standard Sky-Object-Object-Sky dither sequence. Observations were obtained in the J-band using the $J\_G5603$ grating and $ZJ\_G0601$ filter, and in the $K_{l}$-band using the $Kl\_G5607$ grating and $HK\_G0603$ filter.

On-source and sky position observations were both obtained with individual exposure times of 550\,s.   
Two sets of observations with six on-source individual exposures were obtained: the first, in the J-band, was centred at 1.25$\mu$m and covered  the spectral range 1.14$\mu$m to 1.36$\mu$m, and the second, in the $K_{l}$-band, was  centred at 2.3$\mu$m and covered the spectral range  2.10$\mu$m to 2.53$\mu$m.

The data reduction procedure included trimming of the images, flat-fielding, sky subtraction, wavelength and spatial distortion calibrations. We also removed the telluric bands and  flux calibrated the frames  by interpolating a black body function to the spectrum of the telluric standard star. These procedures were executed using tasks contained in the {\sc nifs} software package which is part of {\sc gemini iraf} package, as well as generic {\sc iraf} tasks. In order to check our flux calibration, we extracted a nuclear spectrum with the same aperture of a previous spectrum of the galaxy by \citet{rae}. The two spectra are very similar to each other (considering the difference in spectral resolution), with the largest difference in flux being about 5\% at 2.2$\mu$m. The final IFU data cube for each band contains $\sim$4200 spectra, with each spectrum corresponding to an angular coverage of 0\farcs05$\times$0\farcs05, which translates into $\approx$\,15$\times$15\,pc$^2$ at the galaxy and covering the inner 3\arcsec$\times$3\arcsec($\approx$ 900$\times$900\,pc$^2$) of the galaxy.

The full width  at half maximum (FWHM) of the arc lamp lines in the J-band is 1.65 \AA, corresponding in velocity space to 40\,km\,s$^{-1}$, while in the $K_{l}$-band the FWHM of the arc lamp lines is 3.45\AA, corresponding to  45\,km\,s$^{-1}$.
The angular resolution obtained from the FWHM of the spatial profile of the flux distribution of the broad component of the Pa$\beta$ and Br$\gamma$ lines and is 0\farcs21$\pm$0\farcs03 for the J-band and 0\farcs19$\pm$0\farcs03 for the $K_{l}$-band, corresponding to 60\,pc and 55\,pc at the galaxy, respectively.

\begin{figure*}

\includegraphics[width=150mm]{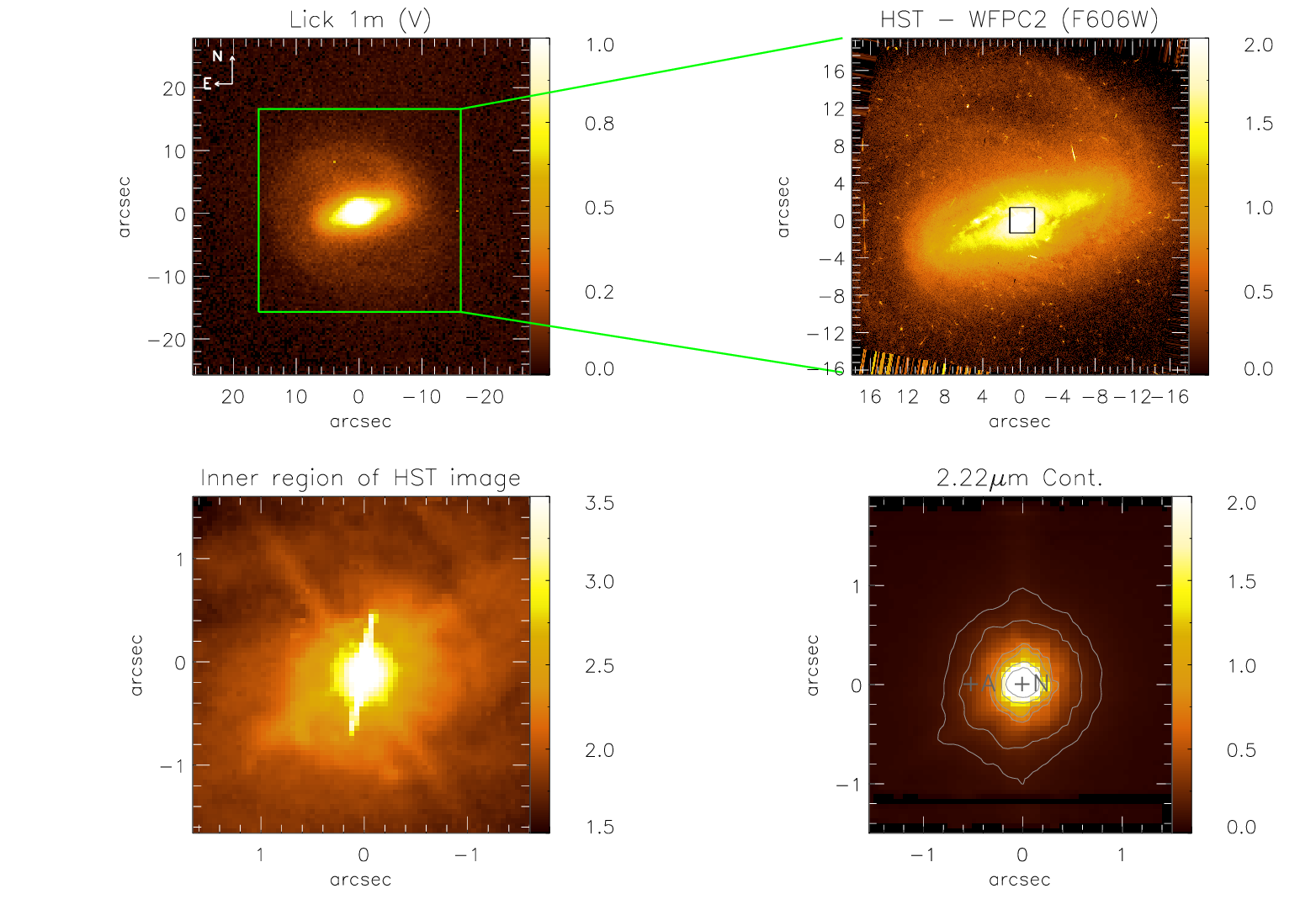}
\caption{Top-left panel: Lick 1m telescope V-band image of Mrk\,766 \citep{hunt} in arbitrary flux units. Top-right panel: HST - WFPC2 continuum image of Mrk\,766 obtained through the filter F606W \citep{ma}.
Bottom-left panel: Zoom of the inner 3\farcs0$\times$3\farcs0 of the HST - WFPC2  image.The color bars for the HST images show the flux in arbitrary units. Bottom-right panel: $2.22\mu$m continuum image obtained from the {\sc nifs} data cube with fluxes shown in units of 10$^{-17}$~erg s$^{-1}$ cm$^{-2}$. The position angle of the major axis of the galaxy is P.A.= 73$^{\circ}$ and the bar is oriented along P.A.= 105$^\circ$. The box in the HST image shows the {\sc nifs} field of view. The labels A and N mark the position where the spectra of Fig.\ref{teste} have been extracted.}
\label{galaxia}
\end{figure*}

\begin{figure*}
\includegraphics[width=150mm]{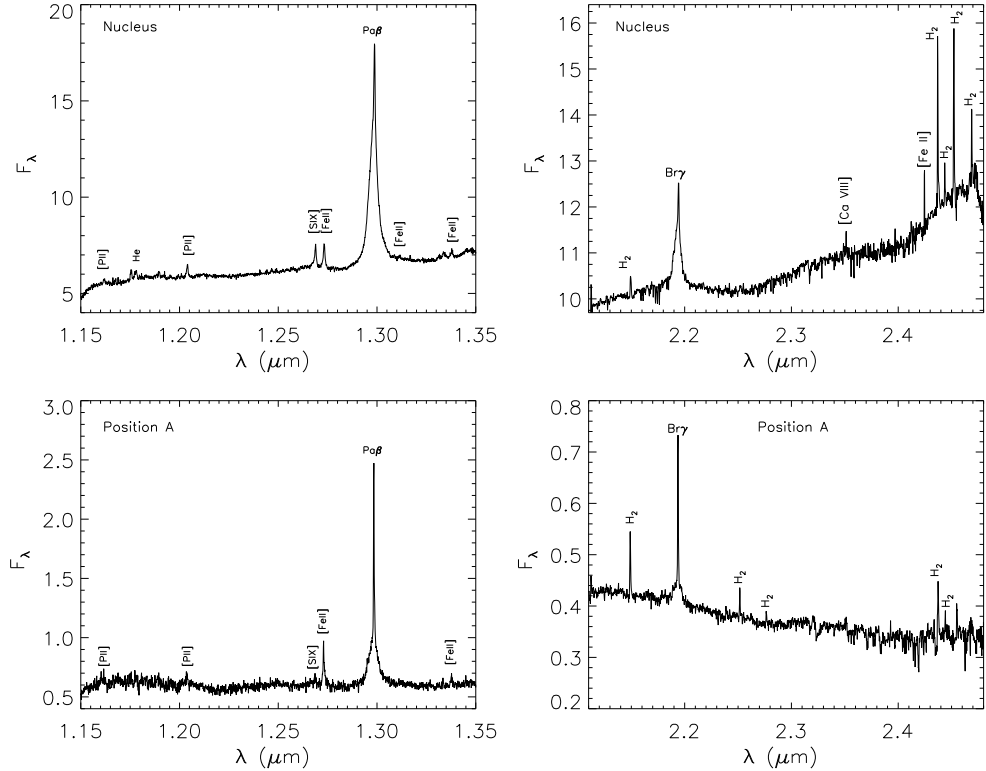}
\caption{Spectra obtained within an 0\farcs25x0\farcs25 aperture centred at the nucleus and at 0\farcs5 east from it (Position A, marked in Fig.\ref{galaxia}). The flux is in $10^{-17}$\,erg\,s$^{-1}$\,cm$^{-2}$\,\AA$^{-1}$ units.}
\label{teste}
\end{figure*}

\section[]{RESULTS}

In the top-left panel of Fig.\ref{galaxia} we present an optical image of Mrk\,766 obtained with the Lick observatory Nickel telescope \citep{hunt}. In the top-right panel we present an optical image of Mrk\,766  obtained with the Hubble Space Telescope (HST) Wide Field Planetary Camera 2 (WFPC2) through the filter F606W \citep{ma}. In the bottom panels we present, to the left, a zoom of the HST image within the field-of-view (FOV) covered by the {\sc nifs} observations and to the right an image obtained from the {\sc nifs} data cube within a continuum window centred at $2.22\mu$m. In Fig.\ref{teste} we present two IFU spectra integrated within a 0\farcs25$\times$0\farcs25 aperture: one at the nucleus and the other  at 0\farcs5 east of it (Position A), chosen randomly with the purpose of just presenting a characteristic extranuclear spectrum. The nucleus was defined to be the location of the peak flux in the continuum.

We list in Table 1 the emission line fluxes  we could measure from these two spectra, which comprise 20 emission lines from the species [P\,{\sc ii}], [Fe\,{\sc ii}], He\,{\sc ii}, H\,{\sc i}, $H_{2}$, [S\,{\sc ix}] and [Ca\,{\sc viii}]. They were measured with the {\it splot} task in {\sc iraf} and the uncertainties were estimated as the standard deviation of the average of 6 measurements.

\subsection{Emission-Line Flux Distributions}

In order to map the flux distributions as well as the centroid velocity and velocity dispersion fields, we used the {\sc profit} routine \citep{rp} to fit the profiles of [P\,{\sc ii}]$\lambda1.1886\mu$m, [S\,{\sc ix}]$\lambda1.2523\mu$m, [Fe\,{\sc ii}]$\lambda1.2570\mu$m, Pa$\beta\lambda1.2822\mu$m, $H_{2}\lambda2.1218\mu$m and Br$\gamma\lambda2.1661\mu$m emission lines at each pixel over the whole FOV. These emission lines were chosen because they have the highest signal-to-noise (S/N) ratios among their species (coronal, ionized and molecular gas). The flux values (as well as those of the central wavelength and width of the profile, see next sections) were obtained by the fit of the profiles using both Gaussian and Gauss-Hermite (GH) series. We found out that the latter gave better fits to most lines, except for the [S\,{\sc ix}] line, for which the GH fits introduced extra wings in some regions where the line was weak. We decided then to adopt the paramenters of the fit obtained from the GH for all lines except for [S IX], for which we adopted the fit with Gaussians. In the case of Pa$\beta$ and Br$\gamma$ we have fitted also a broad component to the line. This was done via a modification of the {\sc profit} routine to fit the broad component and subtract its contribution from the profiles in order  to generate a datacube only with the narrow component. The steps in this procedure were: i ) fit only one Gaussian to the broad component; ii ) subtract it from the spectra where it is present, and iii ) fit the narrow component.

\begin{figure*}
\includegraphics[width=150mm]{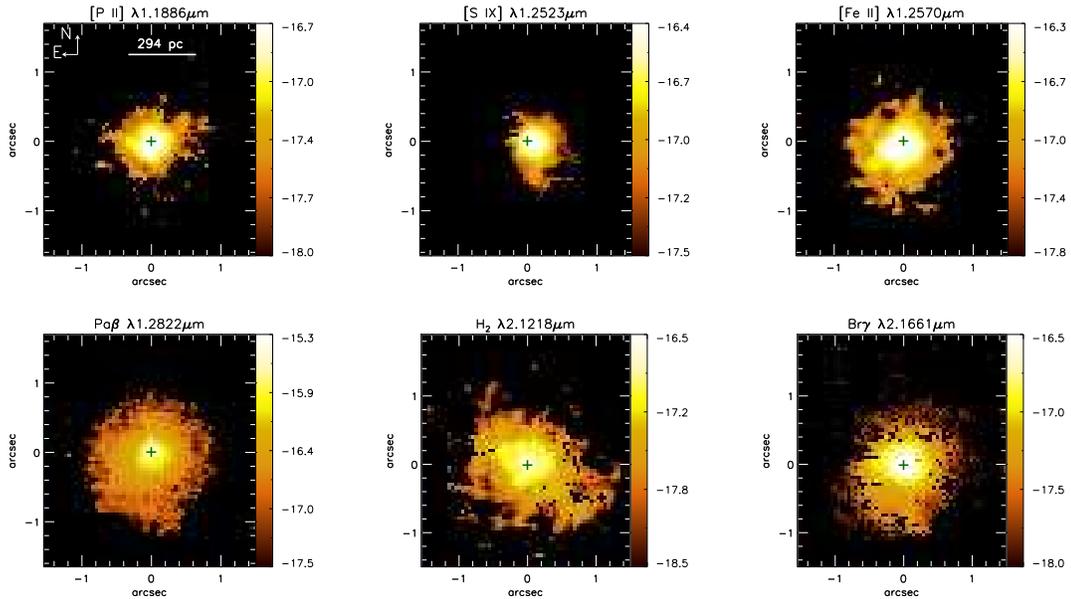}
\caption{Emission-line flux distributions. Flux levels are shown according to the color bar in logarithmic units (erg\,s$^{-1}$\,cm$^{-2}$)}.
\label{flux-distributions}
\end{figure*}

In Fig.\ref{flux-distributions} we present the resulting flux distribution maps, where we have masked out the bad fits by using the chi-square map, which is an output from the {\sc profit} routine. All maps have their peak fluxes at the same position, which also coincides with that of the peak of the continuum: the nucleus. 

The [P\,{\sc ii}] and  [S\,{\sc ix}] flux distributions are the most compact, reaching about 0\farcs5 from the nucleus in all directions in the case of the former, and being more extended to the south-west in the case of the latter. Another coronal line (not shown in the figure), [Ca\,{\sc viii}]$\lambda2.3220\mu$m, also shows a similarly compact flux distribution, indicating that the coronal line region is compact but resolved, extending up to 150 pc from the nucleus, what is a typical radius for this region \citep[e.g.][]{rae2,sbe2,mae,rs}. 

The highest levels of the  [P\,{\sc ii}] flux distribution are more elongated towards the south-east. This elongation is also observed in the [Fe\,{\sc ii}] emission, which reaches 0\farcs8 (240 pc) from the nucleus in that direction. The Pa$\beta$ flux distribution is the most extended in all directions, reaching up to 1$^{\prime\prime}$ from the nucleus. The Br$\gamma$ flux distribution is very similar to that of Pa$\beta$, although noisier due to its lower flux. The H$_2$ flux distribution is somewhat distinct, being elongated from north-east to south-west, thus approximately perpendicular to the elongation of the [Fe\,{\sc ii}] flux distribution, reaching 1\farcs5 (440 pc) from the nucleus towards the south-west.

\subsection{Emission-line ratios}

\begin{figure*}

\includegraphics[width=150mm]{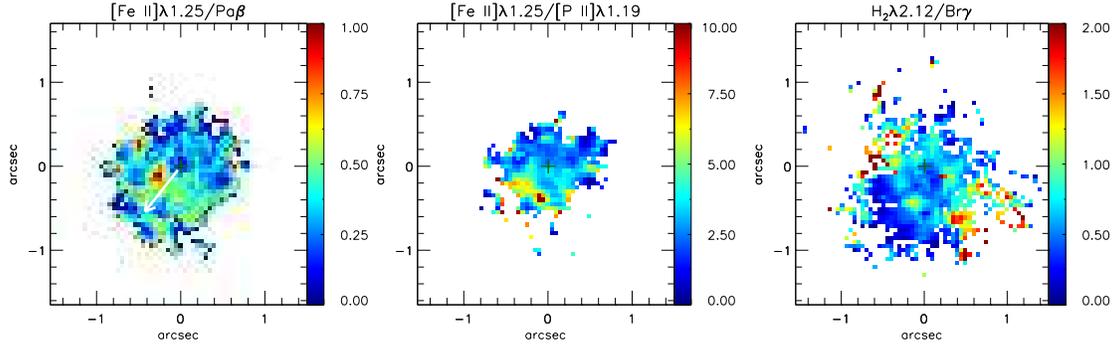}
\caption{Emission-line ratio maps. In the left panel we show the [Fe\,{\sc ii}]$\lambda1.2570\mu$m/$Pa\beta$ ratio map, in the central panel the [Fe\,{\sc ii}]$\lambda1.2570\mu$m/[P\,{\sc ii}]$\lambda1.8861\mu$m ratio map and in the right panel the $H_{2}\lambda2.1218\mu$m/Br$\gamma$ ratio map. The central cross marks the position of the nucleus. The white arrow in [Fe\,{\sc ii}]/Pa$\beta$ panel shows the extent of the radio structure \citep{kk}.}
\label{razoes}
\end{figure*}

In Fig.~\ref{razoes}, we present line-ratio maps obtained from the flux maps, where regions with bad fits were masked out. The average uncertainties in the line-ratio values are $\approx$ 10\%.
In the left panel we present the [Fe\,{\sc ii}]$\lambda\,1.2570\,\mu$m/$Pa\beta$ ratio map, which can be used to investigate the excitation mechanism of [Fe\,{\sc ii}] \citep[e.g.][]{rae,rrp,sbe2,rs}. 
The values of [Fe\,{\sc ii}]/Pa$\beta$ for Mrk\,766 range from 0.2 (most locations) to 1.0 with the highest values -- between 0.6 and 1 -- being observed between 0\farcs2 and 0\farcs6 to the south-east of the nucleus.
Another line ratio that can be used to investigate the [Fe\,{\sc ii}] excitation mechanism is [Fe\,{\sc ii}]$\lambda1.2570\,\mu$m/[P\,{\sc ii}]$\lambda1.8861\mu$m. Values larger than 2 indicate that shocks have passed through the gas destroying the dust grains, releasing the Fe and enhancing its abundance and emission \citep[e.g.][]{ol,sbe2,rsn,rs}. We present this ratio map in the central panel of Fig.~\ref{razoes}. 
The lowest values ($\le$2), are observed to the north and north-west of the nucleus while the highest values of $\approx$ 7.0 are observed in a narrow strip at $\approx$\,0\farcs6 to the south-east of the nucleus, approximately at the border of the region with the highest values of  [Fe\,{\sc ii}]/Pa$\beta$.

In the right panel of Fig.~\ref{razoes} we present the $H_{2}\lambda2.1218\mu$m/Br$\gamma$ ratio map, which is useful to investigate the excitation of the $H_{2}$ emission line \citep[e.g.][]{rae,rrp,sbe2,re2,re3,re4}. 
In Mrk\,766, the values of this ratio range from 0.2 to 2.0.  The lowest values are observed at the nucleus and in most regions to the south and south-east, except for the region between 0\farcs2 and 0\farcs6 to the south-east where the values increase to $\approx$\,0.8 (where both  [Fe\,{\sc ii}]/Pa$\beta$ and [Fe\,{\sc ii}]/[P\,{\sc ii}] show larger values).  Such increased values are also observed to the north where they reach $\approx$ 1.25. At approximately 1 arcsec to the south-west, the values reach up to $\approx$\,2, although the fit of the lines is not so good and the uncertainty is high there.

 \begin{table*}
 \centering

  \begin{tabular}{@{}llcc@{}}
  \hline
& & Nucleus & Position A  \\
\hline

$\lambda_{vac}$($\mu$m) & ID & F($10^{-16}$ erg s$^{-1}$ cm$^{-2}$) & F($10^{-16}$ erg s$^{-1}$ cm$^{-2}$)  \\
\hline
1.18861 & [P\,{\sc ii}]$^1 D_2 -^3 P_2$ & 4.89$\pm$0.2 & $-$ \\
1.25235 & [S\,{\sc ix}]$^3 P_1 -^3 P_2$ & 10.45$\pm$0.6 & 0.39$\pm$0.05 \\
1.25702 & [Fe\,{\sc ii}]$a^4 D_{7/2}-a^6 D_{9/2}$ & 12.98$\pm$0.6 &  2.59$\pm$0.2 \\
1.28216 & H\,{\sc i}Pa$\beta$(narrow) & 34.1$\pm$2  & 6.39$\pm$0.82 \\
1.28216 & H\,{\sc i}Pa$\beta$(broad) &  421.7$\pm$20 & 6.39$\pm$0.82 \\
1.32092 & [Fe\,{\sc ii}]$a^4 D_{7/2}-a^6 D_{7/2}$ & 1.28$\pm$0.1 & 0.51$\pm$0.03 \\
2.12183 & $H_{2}$ 1-0S(1) & 5.66$\pm$0.03 & 1.21$\pm$0.17 \\
2.15420 & $H_{2}$ 2-1S(2)& $-$ & $-$ \\
2.16612 & H\,{\sc i}Br$\gamma$(narrow) & 8.46$\pm$0.3  & 2.21$\pm$0.45 \\
2.16612 & H\,{\sc i}Br$\gamma$(broad)  & 90.0$\pm$4 & 2.21$\pm$0.45 \\
2.18911 & He\,{\sc ii} 10-7 & 1.96$\pm$0.12 & $-$ \\
2.22344 & $H_{2}$ 1-0 S(0) & $-$ & 0.41$\pm$0.09 \\
2.24776 & $H_{2}$ 2-1 S(1) & $-$ & 0.25$\pm$0.02 \\
2.32204 & [Ca\,{\sc viii}]$^2 P^{0}_{3/2}-^2 P^{0}_{1/2}$ & 1.56$\pm$0.80 & $-$ \\
2.36760 & [Fe\,{\sc ii}]$a^{4}G_{9/2}-a^{4}H_{9/2}$ & 52.2$\pm$0.83 & $-$ \\
2.39396 & [Fe\,{\sc ii}]$b^{4}D_{7/2}-a^{2}F_{7/2}$ & 10.6$\pm$1.76 & $-$ \\
2.40847 & $H_{2}$ 1-0 Q(1) & 23.70$\pm$2.24 & 1.02$\pm$0.19 \\
2.41367 & $H_{2}$ 1-0 Q(2) & 22.6$\pm$1.30 & 0.32$\pm$0.02 \\
2.42180 & $H_{2}$ 1-0 Q(3) & 36.70$\pm$2.01 & $-$ \\ 
2.43697 & $H_{2}$ 1-0 Q(4) & 75.7$\pm$1.12 & $-$ \\

\hline
 \end{tabular}
\caption{Measured emission-line fluxes (in units of 10$^{-16}$ erg s$^{-1}$ cm$^{-2}$) for the two positions identified  in Fig.\,\ref{galaxia}.}

\end{table*}

\subsection{Gas Kinematics}

\begin{figure*}
\includegraphics[width=150mm]{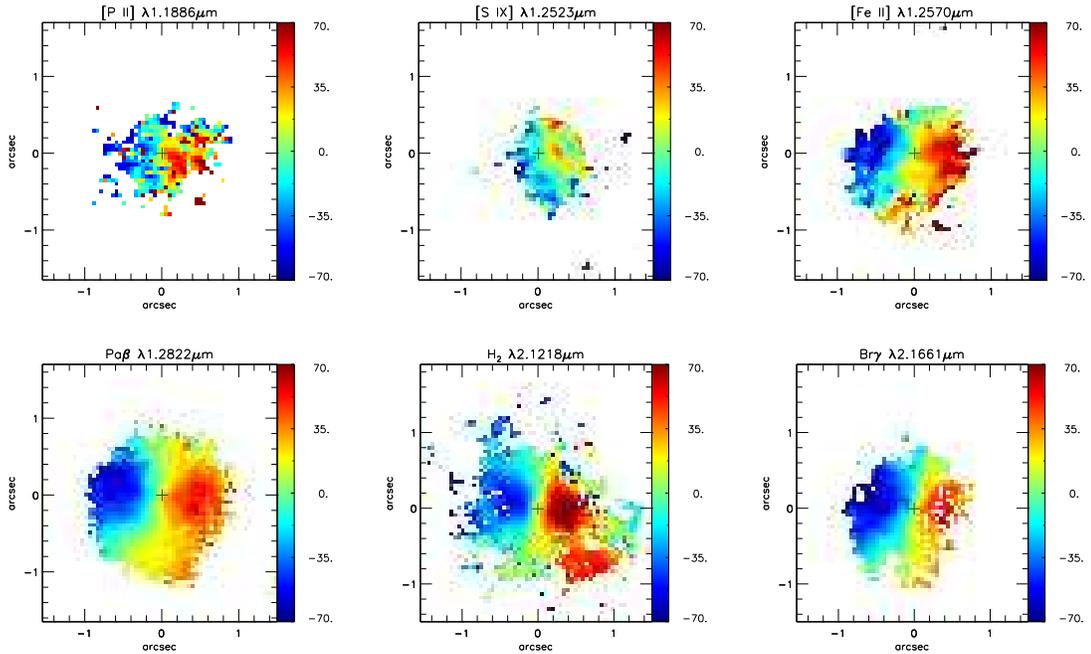}
\caption{Centroid velocity field for the [P\,{\sc ii}]$\lambda1.1886\mu$m (top left), [S\,{\sc ix}]$\lambda1.2523\mu$m (top middle), [Fe\,{\sc ii}]$\lambda 1.2570\mu$m (top right), Pa$\beta$ (bottom left), $H_{2}\lambda2.1218\mu$m (bottom middle) and Br$\gamma$ (bottom right) emitting gas. The central cross marks the position of the nucleus. The color bar shows the velocities in units of km\,s$^{-1}$.}
\label{velocity}
\end{figure*}

The {\sc profit} routine (Riffel 2010) that we have used to obtain the flux of the emission lines, provide also the centroid velocity (V), velocity dispersion ($\sigma$) and higher order Gauss-Hermite moments ($h_3$ and $h_4$), which have been used to map the gas kinematics.
In Fig.~\ref{velocity} we present the centroid velocity fields after subtraction of the heliocentric systemic velocity of 3853$\pm$17 km s$^{-1}$, which was obtained through a model fitted to the Pa$\beta$ velocity field, as discussed in Sec. 4.3. The uncertainties in the velocity maps range from 5 to 20 km s$^{-1}$ depending on the S/N ratio of the spectra (which decrease from the center towards the border of the mapped region). 
The white regions in the figures represent locations where the S/N was not high enough to allow the fitting of the line profiles.
All velocity fields show blueshifts to the east (left in the figures) and redshifts to the west, with the line of nodes oriented at a position angle of approximately $80^\circ$ (see Sec. 4.3), with the isovelocity lines showing an approximate ``spider diagram''  characteristic of rotation. 

Fig. 6 shows the velocity dispersion maps corresponding to the centroid velocity maps of Fig.~\ref{velocity}. As in the case of the centroid velocities, the uncertainties in the velocity dispersion maps range from 5 to 20 km s$^{-1}$ depending on the S/N ratio of the spectra. The white regions in the figures represent locations where the S/N was not high enough to allow the fitting of the line profiles. The [Fe\,{\sc ii}] $\sigma$ map shows the highest values of up to 
150 km\,s$^{-1}$ to the south-east of the nucleus and lowest values, down to 75 km\,s$^{-1}$, to the north-west. The [P\,{\sc ii}] $\sigma$ map has medium values with soft deviations. The Pa$\beta$ $\sigma$ map shows high values at the nucleus and
also 0$\farcs$4 to the south and 0$\farcs$6 to the north and lower values to the east, south and west of this central region. The higher values at the nuclear region may be due to residual contamination from the broad component of the line. The $H_2$ emitting 
gas presents the lowest $\sigma$ values, which are smaller than 70 km\,s$^{-1}$ at most locations. We do not show the $h_3$ and $h_4$ maps because their values are low and do not present any systematic behavior.

\begin{figure*}
\includegraphics[width=150mm]{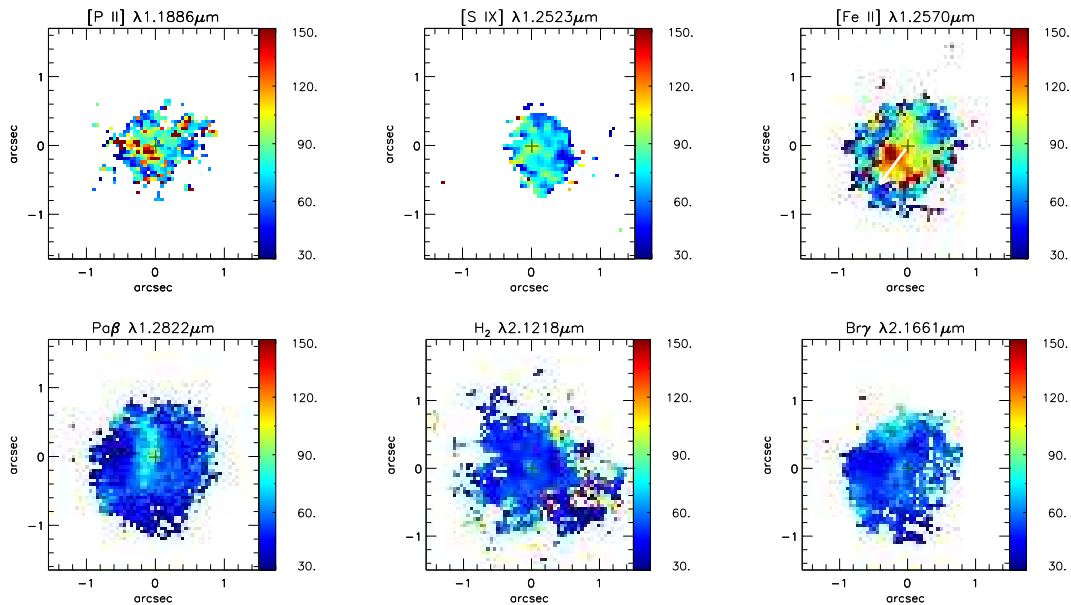}
\caption{$\sigma$ maps for the same emission lines of Fig.\ref{velocity}. The central cross marks the position of the nucleus. The color bars show the $\sigma$ values in units of km\,s$^{-1}$. The white arrow in [Fe\,{\sc ii}] panel shows the extent of the radio structure \citep{kk}.}
\label{disper}
\end{figure*}

\subsection{Channel Maps}

Channel maps along the emission line profiles are shown in Figs. 7, 8, 9 and 10 for the [S\,{\sc ix}], [Fe\,{\sc ii}], Pa$\beta$ and $H_2$ emission lines, respectively. Each panel presents the flux distribution in logarithmic units integrated in velocity bins centred at the velocity shown in the top-left corner of each panel (relative to the systemic velocity of the galaxy). The central cross marks the position of the nucleus. We do not show channel maps for [P\,{\sc ii}] and Br$\gamma$ because the [P\,{\sc ii}] maps are similar to those of [Fe\,{\sc ii}] and those for Br$\gamma$ are similar to those of Pa$\beta$ but noisier.

In Fig. 7, the channel maps along the [S\,{\sc ix}] emission line profile show the flux distributions integrated within velocity bins of 25 km\,s$^{-1}$ (corresponding to one spectral pixel). At the highest velocities the emission is extended 0\farcs5 to the south/south-west, and at the lowest velocities, the [S\,{\sc ix}] is concentrated in the nucleus. 

In Fig. 8, the channel maps along the [Fe\,{\sc ii}] emission-line profile show the flux distributions integrated within velocity bins of 105 km\,s$^{-1}$ (corresponding to three spectral pixels) for the highest velocities and 50\,km\,s$^{-1}$ for the central panels (corresponding to two spectral pixels).  
All [Fe\,{\sc ii}] channel maps present flux distributions which are elongated towards the south-east, up to$\approx$\,0\farcs9 (270\,pc) from the nucleus. Both the highest blueshifts and highest redshifts, which reach $250$km\,s$^{-1}$, are also observed to the south-east of the nucleus. 

Fig. 9 shows the channel maps for the Pa$\beta$ emitting gas for the same velocity bins as for [Fe\,{\sc ii}]. The highest blueshifts and redshifts are observed mostly at the nucleus, but are probably due to residuals  of a broad component to the line which was fitted and subtracted. The flux distributions are more extended and more symmetrically distributed around the nucleus than those of the [Fe\,{\sc ii}]  channel maps.

Fig. 10 shows the channel maps for the $H_2$ emitting gas, for velocity bins of 30\,km\,s$^{-1}$. The highest blueshifts and redshifts, reaching $\approx$\,130\,km\,s$^{-1}$, are observed to the north-east and south-west of the nucleus respectively, following the line of nodes of the galaxy, as seen in Fig.~\ref{velocity}. For 
zero and positive velocities there is a structure extending from the nucleus to the south-west.

\begin{figure*}
\includegraphics[width=120mm]{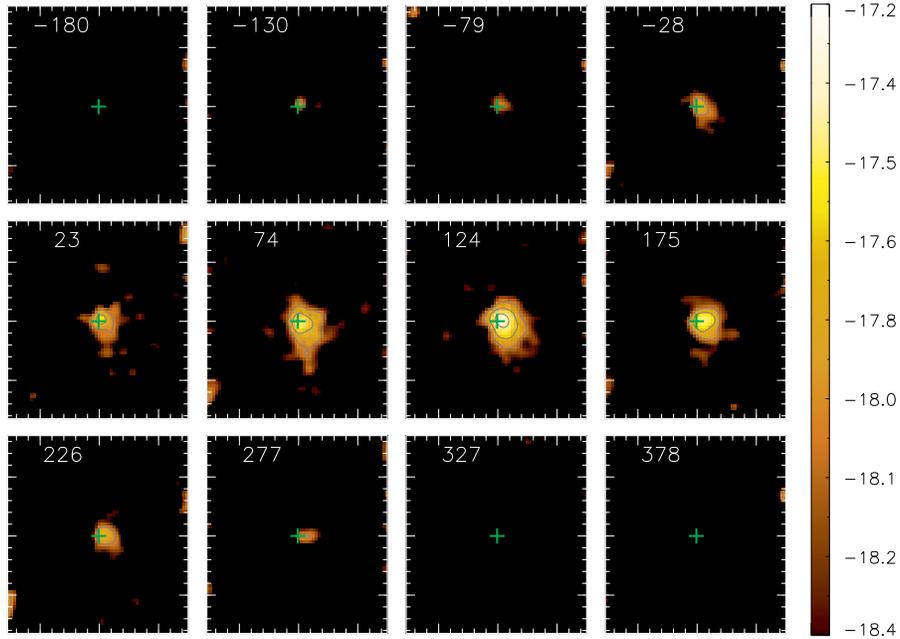}
\caption{Channel maps along the [S\,{\sc ix}] emission line profile, centred at the velocity shown in the upper left corner of each panel in km\,s$^{-1}$. The long tick marks are separated by 1\arcsec and the cross marks the position of the nucleus.}
\label{sliceenxofre}
\end{figure*}

\begin{figure*}
\includegraphics[width=120mm]{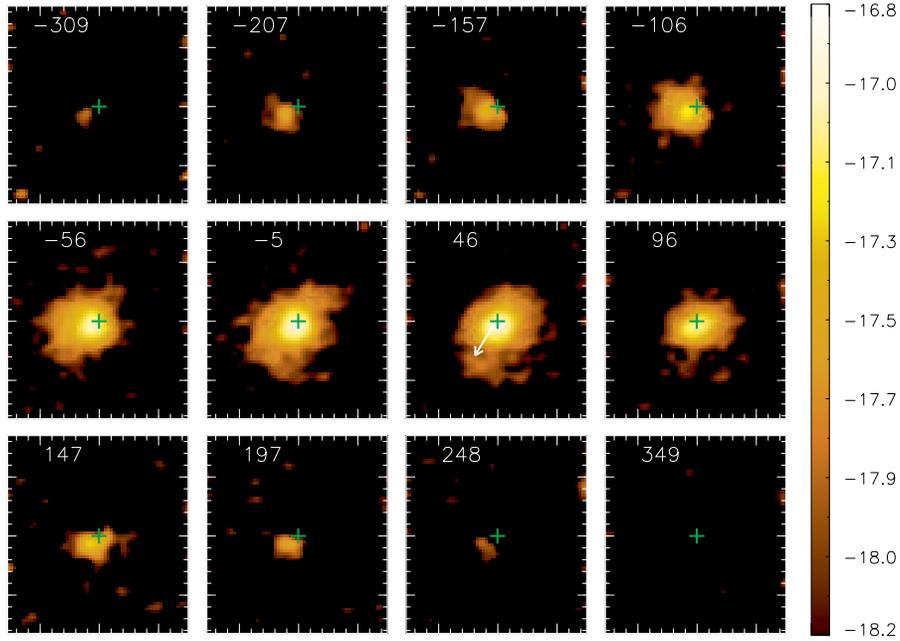}
\caption{Channel maps along [Fe\,{\sc ii}] emission line profile. Description as in Fig.\,\ref{sliceenxofre}, with a white arrow showing the extent of the radio structure \citep{kk}.}
\label{sliceferro}
\end{figure*}

\begin{figure*}
\includegraphics[width=120mm]{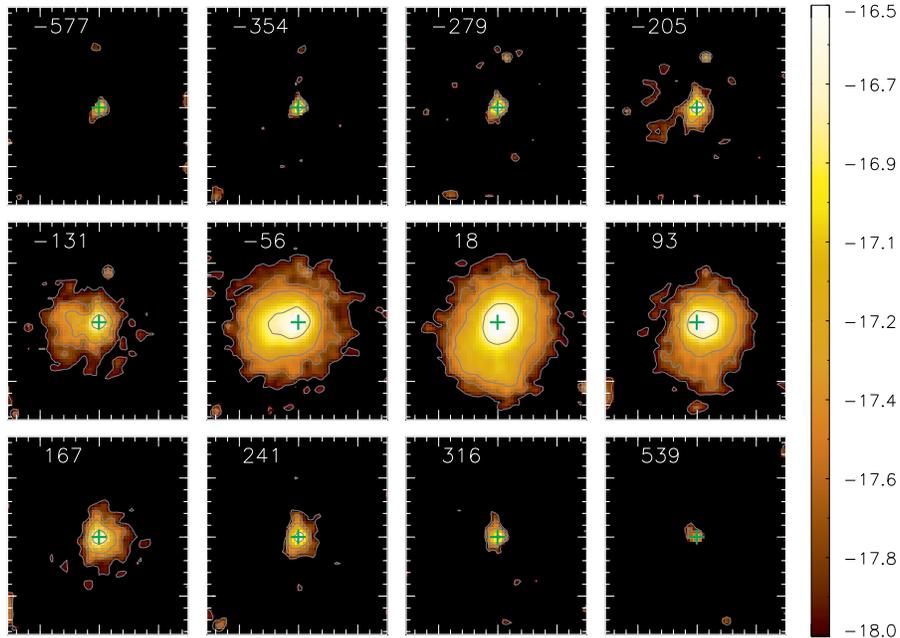}
\caption{Channel maps along $Pa\beta$ emission line profile. Description as in Fig.\,\ref{sliceenxofre}.}
\label{slicepab}
\end{figure*}

\begin{figure*}
\includegraphics[width=120mm]{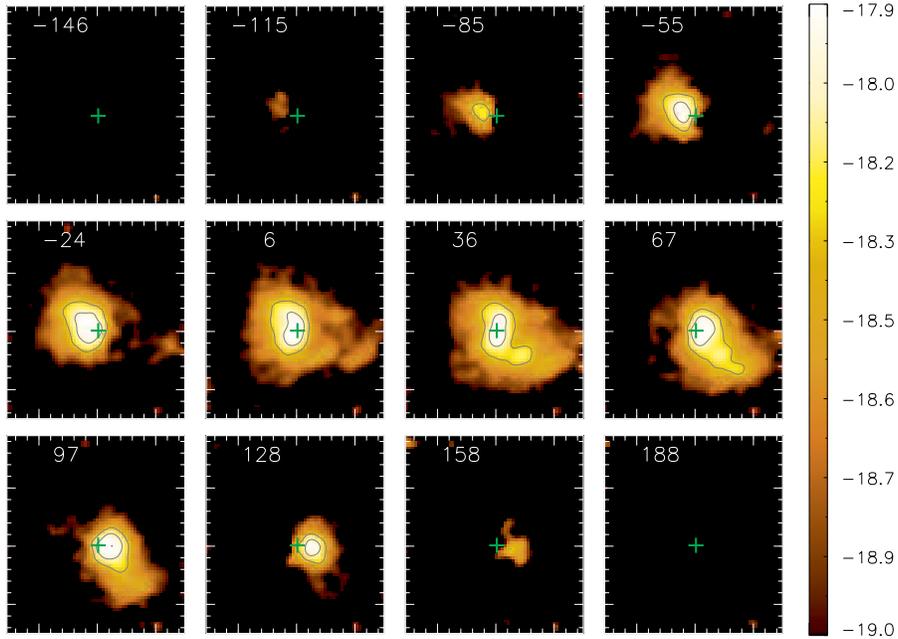}
\caption{Channel maps along $H_2$ emission line profile. Description as in Fig.\,\ref{sliceenxofre}.} 
\label{sliceh2}
\end{figure*}

\section{Discussion}

\subsection{Gaseous Excitation}

\subsubsection{Diagnostic Diagram}

\begin{figure}
\includegraphics[scale=1]{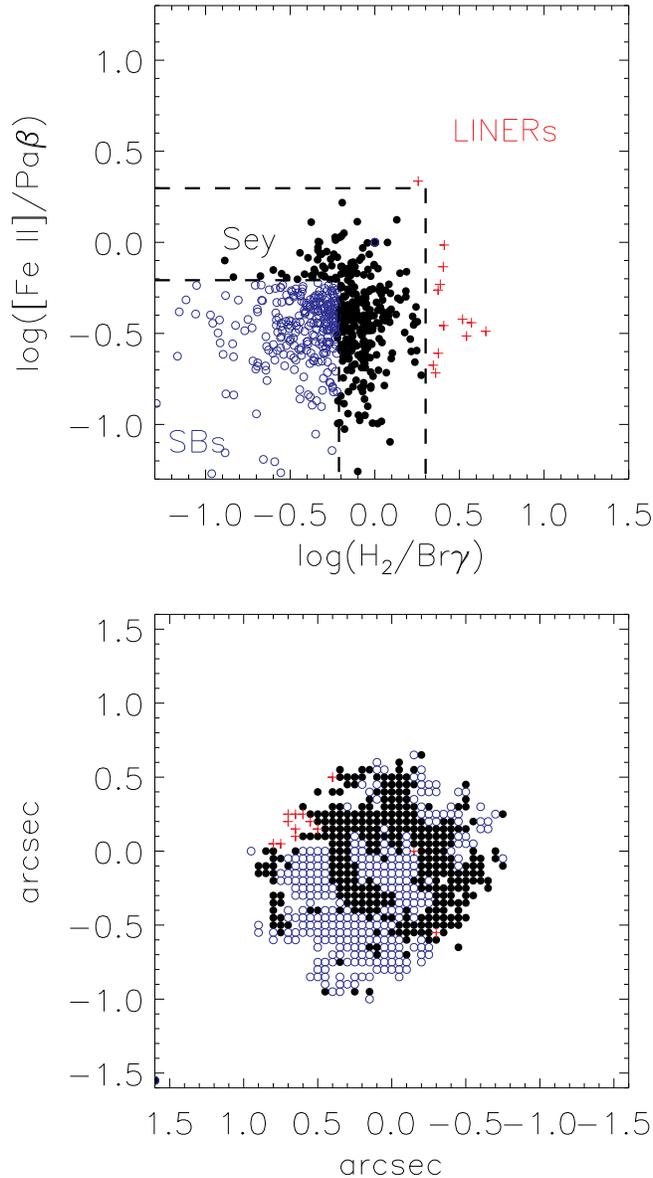}
\caption{Top panel: [Fe\,{\sc ii}]$\lambda$1.25$\mu$m/Pa$\beta$ versus H$_2$$\lambda$2.12$\mu$m/Br$\gamma$ line-ratio diagnostic diagram. The dashed lines delimit regions with ratios typical of Starbursts (blue open circles), Seyferts (black filled circles) and LINERS (red crosses). Bottom panel: spatial position of each point from the diagnostic diagram.}
\label{diagnostico}
\end{figure}

In order to further map the excitation of the circum-nuclear line-emitting region we constructed a spectral diagnostic diagram with the ratios [Fe\,{\sc ii}]$\lambda1.2570\mu$m$/Pa\beta$ vs. $H_{2}\lambda2.1218\mu$m/Br$\gamma$ \citep{lar,rae,rrp,rsn}, shown in Fig\,\ref{diagnostico}. Typical values for the nuclei of Seyfert galaxies range between 0.6 and 2.0 for both ratios \citep{rrp}, while for Starbursts the values are smaller than 0.6 and for LINERs the values are larger than 2, as shown in the top panel of Fig.\,\ref{diagnostico}. In this figure, black filled circles represent Seyfert ratios, blue open circles, Starbursts ratios and red crosses represent ratios of low-ionization nuclear emission-line regions (LINERs). Most ratios present Starburst and Seyfert values, with a few LINER values. The locations from where the distinct line ratios originate are shown in the bottom panel of Fig.\,\ref{diagnostico}. Seyfert ratios are found at the nucleus, in most regions to the north and between 0\farcs2 and 0\farcs6 to the south-east. In a few locations to the north of the nucleus, between the nucleus and the region of enhanced ratios (at 0\farcs2--0\farcs6) and beyond this region, Starburst ratios are found. LINER ratios are only found at the south-west border of the mapped region. 

\subsubsection{The $H_2$ emission}

The excitation of warm H$_2$ has been the subject of many previous studies \citep[e.g.][]{bl,hm,sbe0,rk,rae,rrpa,rrp,da,re2,re3,re4,rsn,sbe2,gu,mrk79,ma2013}. Summarizing, these studies have shown that the $H_2$ emission lines can be excited by two mechanisms: (i) fluorescent excitation through absorption of soft-UV photons (912-1108 \AA) in the Lyman and Werner bands \citep{bl} and (ii) collisional excitation due to heating of the gas by shocks, in the interaction of a radio jet with the interstellar medium \citep{hm} or heating by X-rays from the central AGN \citep{mht}. 
The second mechanism is usually referred to as thermal process since it involves the local heating of the emitting gas, while the first is usually called a non-thermal process. Previous studies have verified that non-thermal processes are not important for most galaxies studied so far \citep[e.g.][]{rae,rrp,rsn}.

In the case of Mrk\,766, the H$_2$/Br$\gamma$ ratio (Fig.~\ref{razoes}) is larger than 0.6 to the north-east of the nucleus and in the arc-shaped region between $0\farcs2$ and $0\farcs5$ to the south-east
supporting Seyfert excitation there. The origin of the H$_2$ excitation could be fluorescence or thermal excitation. One possible evidence for fluorescence (a non-thermal process) is a ratio between the H$_2$ lines 2.24$\mu$m/2.12$\mu$m higher than 0.6 \citep{sbe2}. We could measure this ratio at position A, where the value is $\sim$0.2, favoring thermal excitation. At a number of other positions, the 2.24$\mu$m line is fainter, but wherever it could be measured, the line ratio is smaller than 0.2. This line ratio thus seem to favor thermal excitation due to heating by X-rays or shocks from a radio jet (which seems to be present to the SE) in the region with ``Seyfert excitation", although there is no clear signature of shocks such as an increase in velocity dispersion as observed in the [FeII] emission.

In the remaining regions, the H$_2$/Br$\gamma$ ratio is smaller than 0.6, supporting starburst excitation via heating from shocks in SNe winds and/or by UV radiation from young stars. The presence of starbursts in the nuclear region is in agreement with the results of \citet{rav}, who reported the observation of the PAH 3.3$\mu$m feature in the infrared spectrum of Mrk\,766 within the inner 150 pc, suggesting the presence of recent star formation there. The mixed Seyfert and Starburst excitation is also seen in the diagnostic diagram of Fig.~\ref{diagnostico}.



\subsubsection{The [Fe\,{\sc ii}] emission}

Using the [Fe\,{\sc ii}]$\lambda1.2570\mu$m$/Pa\beta$ and [Fe\,{\sc ii}]$\lambda1.2570\mu$m$/$[P\,{\sc ii}]$\lambda1.8861\mu$m line-ratio maps shown in Fig.\,\ref{razoes}, we can investigate the excitation mechanism of [Fe\,{\sc ii}]. The first ratio [Fe\,{\sc ii}]$\lambda1.2570\mu$m$/Pa\beta$ is controlled
by the ratio between the volumes of partially to fully ionized gas regions, as the [Fe\,{\sc ii}] emission is excited in partially ionized gas regions. In AGNs, such regions can be created by X-ray \citep[e.g.][]{sim} and/or shock \citep[e.g.][]{fow} heating of the gas. 
For Starburst galaxies, [Fe\,{\sc ii}]/$Pa\beta\le0.6$ and for supernovae for which shocks are the main excitation mechanism, this ratio is larger than 2 \citep{rae,rrp}. 

The values of [Fe\,{\sc ii}]/$Pa\beta$ range from $\approx$\,0.2 to the north-west to $\approx$\,1.0 in the arc-shaped region between 0\farcs2 and 0\farcs6 to the south-east of the nucleus. 
\citet{kk} have obtained a 3.6cm radio image of Mrk\,766 and found an extended emission to the SE, at the location of the arc-shaped region where there is an enhancement of the [Fe\,{\sc ii}]/Pa$\beta$ ratio. The variation of this line ratio, and its correlation with the radio structure suggest that excitation by shocks from the radio jet is indeed important at this location. On the other hand, we can not rule out the possible contribution from supernovae as well.

The above conclusion is also supported by the [Fe\,{\sc ii}]$\lambda1.2570\mu$m/[P\,{\sc ii}]$\lambda1.1886\mu$m line-ratio map (central panel of Fig.\,\ref{razoes}). These two lines have similar ionization temperatures, and their parent ions have similar ionization potentials and relative  recombination coefficients. Values larger than 2 indicate that the shocks have passed through the gas destroying the dust grains, releasing the Fe and enhancing its abundance and thus emission \citep{ol,sbe2,rsn}. For supernovae remnants, where shocks are the dominant excitation mechanism, [Fe\,{\sc ii}]/[P\,{\sc ii}] is typically higher than 20 \citep{ol}. For Mrk\,766, to the south-east of the nucleus, where there is the radio structure, [Fe\,{\sc ii}]/[P\,{\sc ii}] values reach $\approx$\,10 (Fig.\,\ref{razoes}), suggesting that shocks are indeed important  in agreement with the highest values obtained for the [Fe\,{\sc ii}]/$Pa\beta$ at the same locations. In other regions, typical values are [Fe\,{\sc ii}]/[P\,{\sc ii}]$\le$ 2, indicating almost no contribution from shocks.

Finally, the diagnostic diagram of Fig.\,\ref{diagnostico} confirms Seyfert excitation in the nucleus and in the south-east arc and regions surrounding the nucleus and non-Seyfert values in the other regions. And the low [Fe\,{\sc ii}]/[P\,{\sc ii}] ratios in these other regions suggest also that SNe winds should not be important, favoring ionization by young stars instead.




\subsection{Mass of ionized and molecular gas}

The mass of ionized gas in the inner 900\,$\times$\,900\,pc$^2$ of the galaxy can be estimated using \citep[e.g.][]{re3,sbe2,scoville}:

\begin{equation}
M_{HII} \approx 3 \times 10^{17}\left(\frac{F_{Br\gamma}}{erg\,s^{-1}cm^{-2}}\right) \left(\frac{D}{Mpc}\right)^2 [M_\odot] ,
\end{equation}
where $F_{Br\gamma}$ is the integrated flux for the $Br\gamma$ emission line and $D$ is the distance to Mrk\,766. We have assumed an electron temperature $T=10^4 K$ and electron density $N_e=100\,$cm$^{-3}$ \citep{osterbrock}. 

The mass of warm molecular gas can be obtained using \citep{scoville}: 

\begin{equation}
M_{H_2} \approx 5.0776 \times 10^{13} \left(\frac{F_{H_{2}\lambda2.1218}}{erg\,s^{-1}cm^{-2}}\right) \left(\frac{D}{Mpc}\right)^2 [M_\odot], 
\end{equation}
where $F_{H_{2}\lambda2.1218}$ is the integrated flux for the $H_2\lambda2.1218\mu$m emission line and  we have used the vibrational temperature $T=2000 K$ \citep{re3,rsn,sbe2}.

We used the Br$\gamma$/Pa$\beta$ line ratio in order to estimate the effect of the reddening in the observed fluxes for these lines. We constructed a reddening map using the Pa$\beta$/Br$\gamma$ line ratio. The resulting map is very noisy with an additional uncertainty relative to other line ratios because the lines are in different spectral bands (K and J). The E(B-V) values are also mostly very small. Thus, instead of using this map to correct the whole Br$\gamma$ flux distribution for reddening -- as we would introduce too much noise -- we have estimated an average value for $E(B-V)=0.3\pm0.1$ using the the integrated fluxes for Br$\gamma$ and Pa$\beta$ emission lines over the whole field of view, following \citet{sbe2} and adopting the extinction law of \citet{ccm}.

Adopting this $E(B-V)$ value, the fluxes for the emission lines in the K band increase by about 10\%. The effect of the reddening is negligible for the line ratios of Fig.~\ref{razoes}, since the lines  are from the same band and the reddening has no effect on the discussion of the gas excitation presented above. On the other hand, its effect is not negligible for the estimate of the ionized and molecular gas masses, which have thus been corrected. Integrating over the whole IFU field, we obtain the following reddening-corrected values: $F_{Br\gamma}\approx$ 6.82$\pm0.35\times10^{-15}$ erg$\,s^{-1}$cm$^{-2}$ and $F_{H_{2\lambda2.1218}}\approx$7.3$\pm0.37\times10^{-15}$erg$\,s^{-1}$cm$^{-2}$. The resulting masses are  $M_{HII}\approx$ 7.6$\pm0.4\times10^{6}\,M_\odot$ and $M_{H_{2}}\approx$ 1.32$\pm0.07\times10^{3}$ $M_\odot$.

The above values are similar to those we have obtained in previous studies, which are in the range $0.1\times10^{6}$ M$_\odot\le$ M$_{HII}\le$1.7$\times10^{6}$ M$_\odot$ and  66M$_\odot\le$ M$_{H_2}\le$3300 M$_\odot$, respectively.

The mass of molecular gas is thus 10$^3$ times smaller than that of the ionized gas but, as discussed in \citet{sbe2}, this H$_2$ mass represents only that of warm
gas emitting in the near-IR. The total mass of molecular gas is dominated by the cold gas, and the usual proxy to estimate the cold H$_2$ mass has been the CO emission. A number of studies have derived the ratio between the cold and warm H$_2$ gas masses by comparing the masses obtained using the CO and near-IR emission. \citet{dale05} obtained ratios in the range 10$^5$--10$^7$; using a larger sample of 16 luminous and ultraluminous infrared galaxies, \citet{ms2006} derived a ratio M$_{cold}$/M$_{warm}$ = 1-5$\times\,10^6$. More recently, \citet{ma2013} compiled from the literature values of Mcold derived from CO observations and H$_2$2.12$\mu$m luminosities for a larger number of galaxies, covering a wider range of luminosities, morphological and nuclear activity types. From that, an estimate of the cold H2 gas mass can be obtained from



\begin{equation}
M_{H_2\,{\rm cold}} \approx 1174\,\left(\frac{L_{H_2\lambda2.1218}}{L_{\odot}}\right),
\end{equation}
where $L_{H_2\lambda2.1218}$ is the luminosity of the H$_2$2.12$\mu$m line. The resulting mass value is M$_{H_2\,cold}$ $\approx$ 9.8$\times$10$^{8}\,M_{\odot}$.

\subsection{Gaseous Kinematics}

All the velocity fields shown in Fig.~\ref{velocity} suggest rotation in the inner 450\,pc of Mrk\,766. In order to obtain the systemic velocity, orientation of the line of nodes and an estimate for the enclosed mass, we fitted a model of circular orbits in a plane to the $Pa\beta$ and H$_{2}\lambda2.1218\mu$m velocity fields. The expression for the circular velocity is given by \citep{ba,re2,rs}:
\begin{equation}
 V_{r}=V_{s}+\sqrt{\frac{R^2GM}{(R^2+A^2)^{3/2}}}\frac{\sin(i)\cos(\Psi-\Psi_{0})}{\left(\cos^2(\Psi-\Psi_{0})+\frac{\sin^2(\Psi-\Psi_{0})}{\cos^2(i)}\right)^{3/4}}
\end{equation}
where $R$ is the projected distance from the nucleus in the plane of the sky, $\Psi$ is the corresponding position angle, $M$ is the mass inside $R$, $G$ is the Newtow's gravitational constant, $V_s$ is the systemic velocity, $i$ is the inclination of the disc ($i=0$ for a face-on disc), $\Psi_0$ is the position angle of the line of nodes and $A$ is a scale length projected in the plane of the sky. 

\begin{figure*}
\includegraphics[scale=0.9]{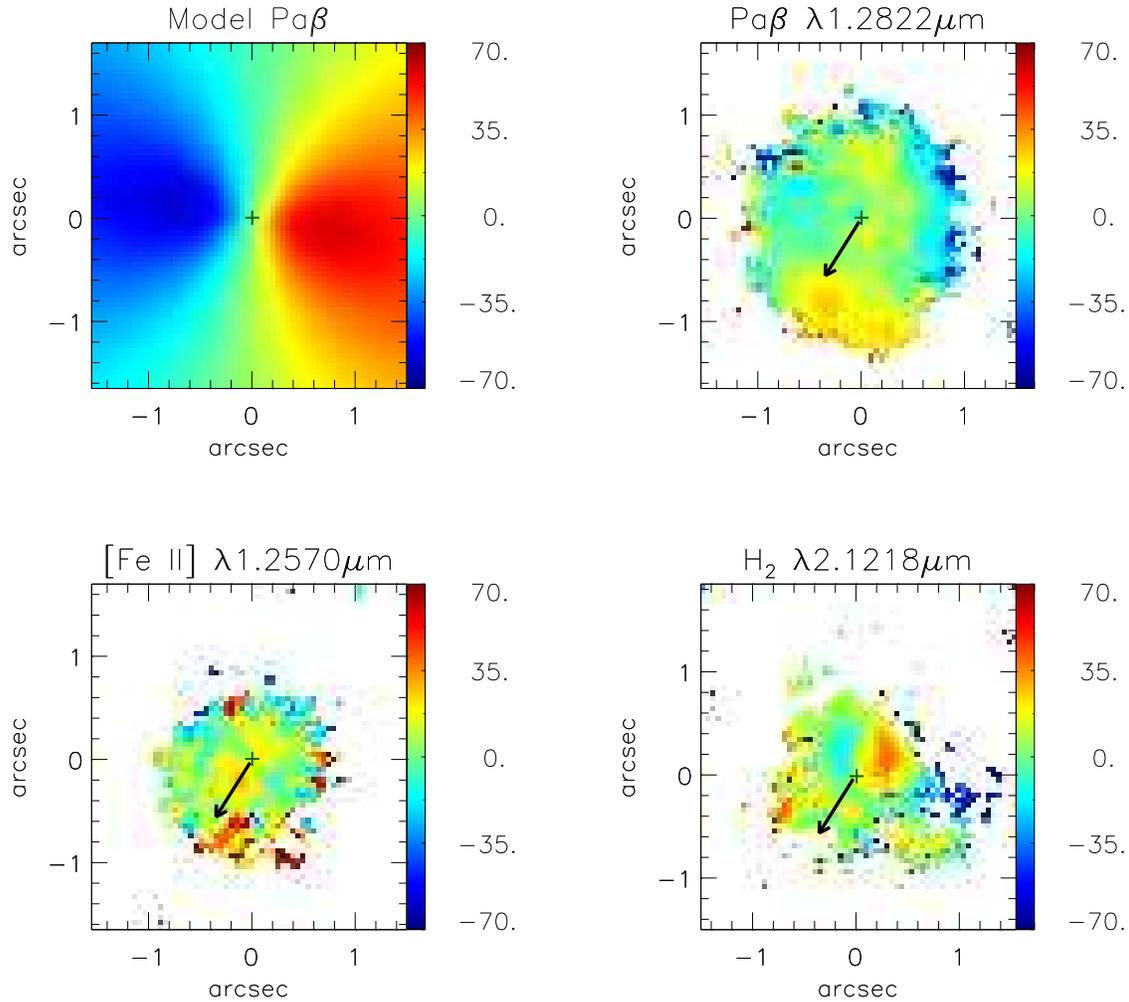}
\caption{Rotating disc model fitted to the Pa$\beta$ velocity field, together with the residuals of its subtraction from the observed velocity fields of Pa$\beta$, [Fe\,{\sc ii}]$\lambda\,1.2570\mu$m and $H_2\lambda\,2.1218\mu$m.The black arrows show the extent of the radio structure \citep{kk}.}
\label{residuals1}
\end{figure*}

\begin{figure*}
\includegraphics[scale=0.9]{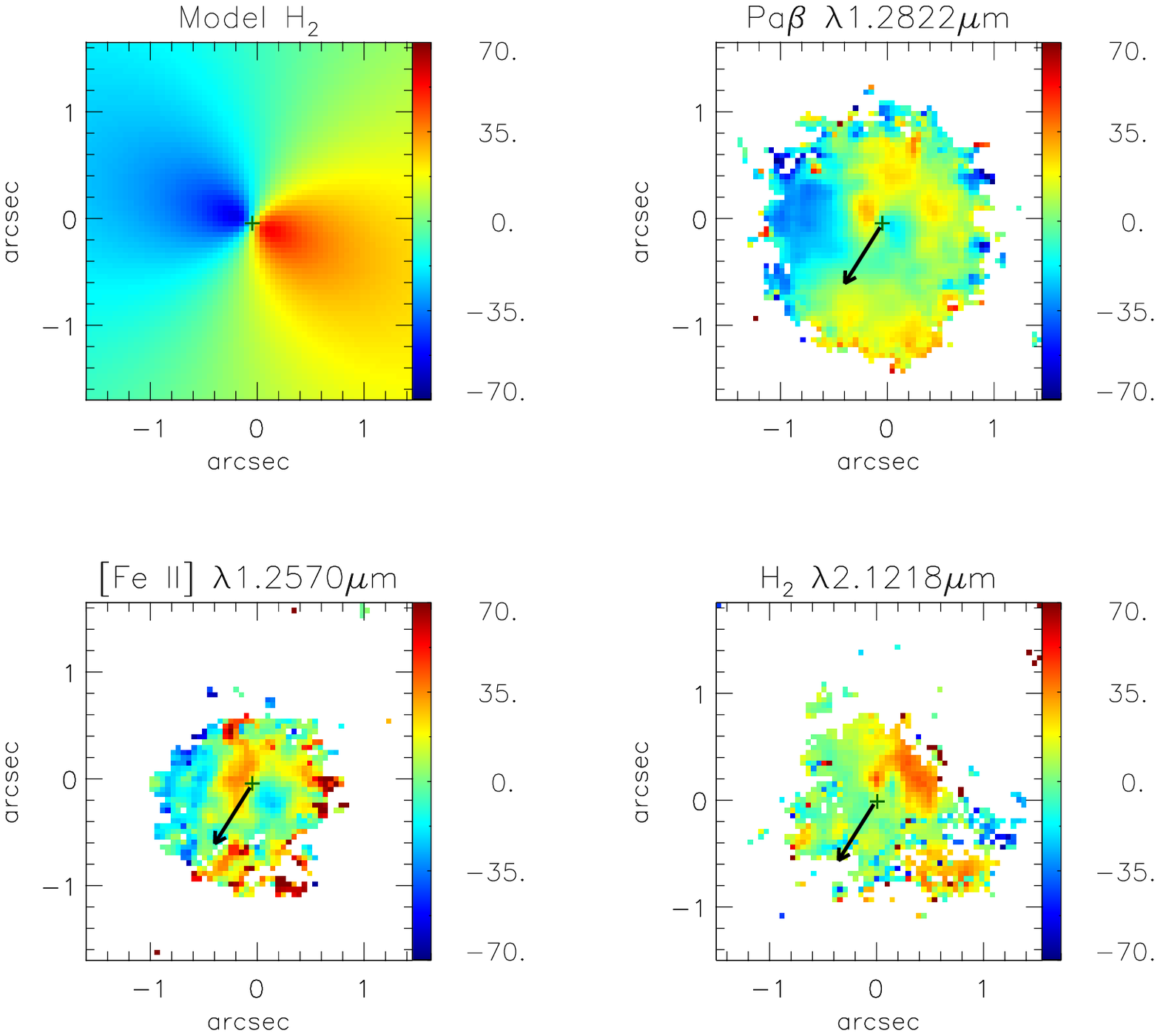}
\caption{Rotating disc model fitted to the H$_{2}\lambda2.1218\mu$m velocity field, together with the residuals of its subtraction from the observed velocity fields of Pa$\beta$, [Fe\,{\sc ii}]$\lambda\,1.2570\mu$m and $H_2\lambda\,2.1218\mu$m. The black arrows show the extent of the radio structure \citep{kk}.}
\label{residuals2}
\end{figure*}

The location of the kinematical center was not allowed to vary, being fixed to the position of the peak of the continuum. The equation above contains five free parameters, which can be determined by fitting the model to the observations. This was done using the Levenberg-Marquardt least-squares fitting algorithm, 
in which initial guesses are given for the free parameters. The best fit model for Pa$\beta$ is shown in Fig.\,\ref{residuals1} (top-left panel) and the best fit model for H$_{2}\lambda2.1218\mu$m in Fig.\,\ref{residuals2}. In both figures we show the residual maps  (observed velocity field - model) for [Fe\,{\sc ii}] (bottom-left panel), $H_2$ (bottom-right panel) and $Pa\beta$ (top-right panel).

The parameters derived from the fit of the Pa$\beta$ are: the systemic velocity corrected to the heliocentric reference frame $V_s=3853\pm17$\,km\,s$^{-1}$, $\Psi_0=80^{\circ}\pm3.36^{\circ}$, $M=8.72\pm0.63\times10^8\,M_\odot$, $i=30^\circ\pm4^\circ$ and $A=163.4\pm10$ pc. 
We can compare the near-IR line-emitting gas kinematics with results obtained in the optical at larger scales. \citet{gdp96} present long-slit spectroscopy of Mrk~766 at kpc scales with the slit oriented along PA$=55^\circ$. They found that the kinematic of the high-excitation gas (traced by the [O\,{\sc iii}]$\lambda5007$ emission) is more perturbed than that of the low-ionization gas (traced by H$\alpha$ and H$\beta$), showing radial motions consistent with gas outflows from the nucleus. The low-ionization gas seems to be dominated by rotation in the plane of the galaxy with a velocity amplitude of $\sim$130~km\,s$^{-1}$. At distances smaller than 1\farcs5 from the nucleus, the velocity amplitude is $<$50\,km\,s$^{-1}$, which is somewhat smaller than the amplitude that we have derived. This is expected, since the slit used by \citet{gdp96} was not oriented along the major axis of the galaxy. \citet{gdp96} quote a photometric major axis orientation of 105$^\circ$, based on an large scale continuum image at $\lambda$5960\AA. The PA of the line of nodes $\Psi_0$ that we have found is 25$^\circ$ smaller than this value. On the other hand, our $\Psi_0$ is in reasonable agreement with the value listed at the Hyperleda  \citep[$\Psi_0\approx73^{\circ}$ --][]{hyperleda}.  Fig.~\ref{galaxia} shows that Mrk\,766 presents a bar with size of 4.5~kpc. The orientation of the bar is similar 
to that of the photometric major axis considered by \citet{gdp96}. As the bar is broad an luminous and the outer parts of the galaxy are faint, we believe they have mistakenly concluded that the direction of the bar was that of the major axis. The systemic velocity and $i$ are in reasonable agreement with the values listed at the Hyperleda  \citep{hyperleda} and NED databases ($V_s\approx3876$\,km\,s$^{-1}$ and $i\approx36^\circ$ and $M$ and $A$ are similar to values found for other Seyfert galaxies using the same model \citep[e.g.][]{ba}. For the H$_{2}\lambda2.1218\mu$m fit we found a much more compact velocity field than that of Pa$\beta$, with a scale lenght $A=41.1\pm4$ pc. The other parameters were practically the same.   
This signature of a more compact rotating disk in H$_2$ than in Pa$\beta$, is similar to that we have found for Mrk\,1066  \citep{rs2010}, indicating that H$_2$ presents a “colder” and more ordered kinematics. An exception is the region to the south-west, which seems to show a detached kinematics. This region is probably a molecular cloud that is not in the galaxy disc.


The residuals shown in Fig.\,\ref{residuals1} show blueshifts in the borders of the measured field in Pa$\beta$ to the north--north-east of the nucleus which we attribute to poor fits of the lines in this region. More significant are the redshift residuals to the south-south-east, a region where the largest residuals in the [Fe\,{\sc ii}] velocity field are also observed. There is where the enhanced [Fe\,{\sc ii}] velocity dispersion and the radio structure are also located. In addition, in this same region, the [FeII] flux distributions in the channel maps of Fig. 8 show both blueshifts and redshifts, with velocities of up to 250 kms−1. We interpret these results as being due to emission of gas in a one-sided outflow oriented along the position angle $\approx$ 135$^{\circ}$. The observation of both blueshifts and redshifts in the channel maps supports that its axis lies approximately in the plane of the sky.
The main residuals in the H$_2$ velocity field that are not in the borders of the field (where the line fits are poorer) are the redshifts observed to the north--northwest. As this is the near side of the galaxy, we speculate that these residuals could be due to inflows in the plane of the galaxy. 
These residuals are seen along the direction of the bar at PA$\approx -60^\circ$. We speculate that they may be associated to inflows along the bar, as predicted by theoretical models \citep[e.g][]{combes04}  and as measured in a few cases \citep[e.g.][]{mundell99}. In previous studies, we have found inflows along nuclear dusty spirals \citep{sm,re3,mrk79}. Indeed, numerical simulations by \citet{maciejewski04a,maciejewski04b} have shown that if a central SMBH is present, shocks can extend all the way to the vicinity of the SMBH and generate gas inflows consistent with the accretion rates inferred in local AGN.
 Similar residuals are also seen in the Pa$\beta$ and [Fe\,{\sc ii}] residual maps: redshifts to north--northwest, also suggesting inflows. We rule out the possibility of these redshifts being due to a counterpart of the south-east outflow once it is observed in redshift over the far side of the galaxy. A possible counterpart should be in blueshift and behind the near side of the galaxy plane. We do not see such a component; one possibility is that it is hidden by the galaxy plane.

The residuals shown in Fig.\,\ref{residuals2}, after the subtraction of the circular velocity model fitted to the H$_2$ velocity field shows similar residuals to the south--south-east for Pa$\beta$ and [Fe\,{\sc ii}], but show additional residuals in the northern part of the field. In the case of the H$_2$ residuals, redshifts are observed also in the region to the south-west, that we have interpreted as due to a detached cloud, that is probably not in the galaxy plane.

\subsubsection{Mass outflow rate}

With the goal of quantifying the feedback from the AGN in Mrk\,766, we estimate the ionized-gas mass outflow rate through a circular cross section with radius $r=0\farcs25\approx75$\,pc located at a distance of $h=0\farcs4$ from the nucleus to the south-east. This geometry corresponds to a conical outflow with an opening angle of $\approx$ 64$^\circ$, estimated from Fig.\,\ref{sliceferro}. The mass outflow rate can be obtained using:
\begin{equation}
 \dot{M}_{out}=m_p\,N_e\,v\,f\,A
\end{equation}
and the filling factor ($f$) can be obtained from
\begin{equation}
 f=\frac{L_{Pa\beta}}{j_{Pa\beta}\,V}
\end{equation}
where $m_p$ is the proton mass, $N_e$ the electron density, $v_{out}$ is velocity of the outflowing gas and $L_{Pa\beta}$ and $j_{Pa\beta}$ are the luminosity and the emission coefficient of Pa$\beta$ \citep{rs}. 

We have assumed that $N_e=500$\,cm$^{-3}$, $L_{Pa\beta}=$1.43$\times10^{39}$ erg s$^{-1}$, $j_{Pa\beta}=4.07\times10^{-22}$ erg cm$^{-3} $s$^{-1}$ and $v_{out}=\,147$\,km\,s$^{-1}/{\rm sin}\,\theta\approx277$\,km\,s$^{-1}$ where $\theta$ is the angle between the wall of the cone (from where we observe the line-of-sight velocity component of 147\,km\,s$^{-1}$) and the plane of sky. The latter velocity value was obtained directly from the channel maps considering that the structure seen to south-east is due to the emission of the walls of the cone.
As described above, the axis of the cone seems to lie close to the plane of the sky. From the estimated aperture of the cone, we adopt a maximum angle between 
the cone and the plane of the sky of $32^\circ$. Under these assumptions we obtain $f=0.18$ and then $\dot{M}_{out}\approx10\,M_{\odot}$ yr$^{-1}$. The value found here for $\dot{M}_{out}$ is in good agreement with those found in \citet{vc}, which range from 0.1 to 10\,${M}_{\odot}$ yr$^{-1}$, it is of the same order of that obtained by \citet{rs}, of 8\,${M}_\odot$ yr$^{-1}$, and is also within the range of the values found by \citet{msea}, which range from 2.5 to 120 ${M}_\odot$ yr$^{-1}$.

Following \citet{sbe3}, we can use the above mass outflow rate to estimate the kinetic power of the outflow using:
\begin{equation}
 \dot{E}\approx\frac{\dot{M}_{out}}{2}(v^{2}_{out}+\sigma^{2})
\end{equation}
where $v_{out}=v_{obs}/{\rm sin}\theta$ is the velocity of the outflowing gas and $\sigma$ is its velocity dispersion. Using $\sigma\approx100$\,km\,s$^{-1}$ (from Fig.\,\ref{disper}) and $v_{out}=v_{obs}/{\rm sin}\theta=277$\,km\,s$^{-1}$ we obtain $\dot{E}\approx2.9\times10^{41}$\,erg\,s$^{-1}$ which is in good agreement with the values obtained for Seyfert galaxies and compact radio sources \citep{mo}. This value is also similar to that obtained for Mrk\,1157 \citep{rs}, of $\dot{E}\approx5.7\times10^{41}$\,erg\,s$^{-1}$, it is within the range of those found by \citet{msea}, between 0.6 to 50$\times10^{41}$\,erg\,s$^{-1}$.

In order to compare the above value of $\dot{E}$ with the bolometric luminosity, we estimate the latter as 10 times the X-ray luminosity, of 3.5$\times10^{41}$ erg s$^{-1}$ \citep{boller01}, resulting in  $\dot{E}\approx$ 0.08 L$_{Bol}$.

Finally, we can calculate the mass acretion rate to feed the active nucleus from\citep{rs}
\begin{equation}
\dot{m}=\frac{L_{bol}}{c^{2}\eta},
\end{equation}
where, $L_{bol}$ is the nuclear bolometric luminosity, $\eta$ is the efficiency of conversion of the rest mass energy of the accreted material into radiation and $c$ is the light speed. The bolometric luminosity was already estimated as 3.5$\times10^{42}$ erg s$^{-1}$. Assuming $\eta\approx$0.1 which is a typical value for a geometrially thin, optically thick accretion disc \citep{fkr2002}, we obtain an accretion rate of $\dot{m}\approx$\,1.4$\times$10$^{-2}$ M$_{\odot}$\,yr$^{-1}$, which is about three orders of magnitude smaller than the mass outflow rate, a ratio compared with those found in our previous studies.



\section{Conclusions}

We have mapped the gas flux distribution, excitation and kinematics from the inner $\approx$\,450\,pc radius of the Seyfert\,1 galaxy Mrk\,766  using near-IR J-and K-band integral-field spectroscopy at a spatial resolution of $\approx$\,60\,pc (0\farcs20). The main conclusions of this work are:
\begin{itemize}
\item The emission-line flux distributions of molecular hydrogen H$_2$ and low-ionization gas are extended to at least $\approx$\,300\,pc from the nucleus; 

\item The H$_2$ line emission is most extended along PA=70$^\circ$, which is close to the position angle of the line of nodes of the gas kinematics; 

\item The [Fe\,{\sc ii}] emission is most extended approximately along the perpendicular direction to the line of nodes of the gas kinematics;

\item The coronal line [S\,{\sc ix}]  emission is resolved and extends up to  $\approx$\,150\,pc from the nucleus; 

\item The emission-line ratios [Fe\,{\sc ii}]/$Pa\beta$ and $H_{2}$/Br$\gamma$ show a mixture of Starburst and Seyfert type excitation; the Seyfert values dominates at the nucleus, to the north-west and in an arc-shaped region between 0\farcs2 and 0\farcs6 to the south-east where a radio jet has been observed, while Starburst values are present at the nucleus and other regions;

\item The enhancement of the [Fe\,{\sc ii}]/[P\,{\sc ii}] line ratio at the location of the radio jet, as well as the corresponding increase in the [Fe\, {\sc ii}] flux and velocity dispersion support a contribution from shocks to the gas excitation in the arc-shaped region to the south-east; in the remaining regions, the favoured excitation mechanism is UV radiation from young stars;



\item The $H_2$ gas kinematics is dominated by rotation in a compact disc with a velocity amplitude of 140 km s$^{-1}$ and low velocity dispersion (40-60 km s$^{-1}$, consistent with orbital motion in the plane of the galaxy);

\item The kinematics of the ionized gas is also dominated by rotation, but channel maps in [Fe\, {\sc ii}] show in addition an outflowing component to the south-east, with an axis lying close to the plane of the sky reaching velocities of $\sim$ 300 km s$^{-1}$, probably associated with the radio jet.

\item The mass outflow rate in ionized gas is estimated to be $\approx$ 10.7 M$_\odot$ yr$^{-1}$ and the power of the outflow estimated to be $\approx\,0.08\,L_{Bol}$;

\item The mass of ionized gas is $M_{HII}\approx\,7.6\times10^{6}\,M_\odot$ while the mass of the hot molecular gas is $M_{H_{2}}\approx\,1.3\times10^{3}\,M_\odot$ and the estimated cold molecular gas mass is M$_{H_2\,cold}$ $\approx$ 9.8$\times$10$^{8}\,M_{\odot}$.


The distinct flux distributions and kinematics of the H$_2$  and [Fe\,{\sc ii}] emitting gas, with the first more restricted to the plane of the galaxy and in compact rotation and the second related with the radio jet and in outflow are common characteristics of 8 Seyfert galaxies (ESO428-G14, NGC\,4051, NGC\,7582,
NGC\,4151, Mrk\,1066, Mrk\,1157, Mrk\,79 and Mrk\,766) we have studied so far
using similar integral-field observations and 2 others (Circinus and NGC\,2110)
using long-slit observations. These results again suggest -- as those found in previous studies -- that the H$_2$
emission is tracer of the AGN feeding, while the [Fe\,{\sc ii}] is a tracer of its feedback.

\end{itemize}

\section{aknowledgments}

This work is based on observations obtained at the
Gemini Observatory, which is operated by the Association of Universities for Research in Astronomy, Inc., under a cooperative agreement with the NSF on behalf of the Gemini partnership: the National Science Foundation (United States), the Science and Technology Facilities Council (United Kingdom), the National Research Council (Canada), CONICYT (Chile), the Australian Research Council
(Australia), Minist\'erio da Ci\^encia e Tecnologia (Brazil) and south-east CYT (Argentina).   This research has made use of the NASA/IPAC Extragalactic Database (NED) which is operated by the Jet Propulsion Laboratory, California Institute of Technology, under contract with the National Aeronautics and Space Administration. We acknowledge the usage of the HyperLeda database (http://leda.univ-lyon1.fr). This work has been partially supported by the Brazilian institutions CNPq, CAPES and FAPERGS.

\appendix

\label{lastpage}

\end{document}